\documentclass[10pt]{article}
\usepackage{longtable}
\usepackage{tabu}

\usepackage[hyphens]{url}
\usepackage{scrextend}
\usepackage[hidelinks]{hyperref}

\usepackage{graphicx}
\usepackage[ruled,linesnumbered]{algorithm2e}
\usepackage[numbers]{natbib}
\usepackage[singlelinecheck=false]{caption}
\DeclareCaptionLabelFormat{blank}{}
\usepackage[bottom]{footmisc}

\usepackage{booktabs} 
\usepackage{authblk}
\usepackage{adjustbox}
\usepackage{color, colortbl}
\usepackage{subcaption} 
\usepackage{amsthm}
\usepackage{amsmath}

\usepackage{pdflscape}
\usepackage{enumitem}
\usepackage{lineno}
\usepackage{float}
\usepackage{wrapfig}

\definecolor{Gray}{gray}{0.9}
\definecolor{White}{rgb}{1,1,1}
\definecolor{LightCyan}{rgb}{0.88,1,1}
\definecolor{orange}{rgb}{1,0.5,0}

\usepackage[space]{grffile}

\makeatletter
\g@addto@macro{\thm@space@setup}{\thm@headpunct{:}}
\makeatother

\begin{document}
\title{\vspace{-2.0cm}
Zooming Into Video Conferencing Privacy and Security Threats}

\author[1]{Dima Kagan\thanks{kagandi@bgu.ac.il}}
\author[1]{Galit Fuhrmann Alpert\thanks{fuhrmann@bgu.ac.il}}
\author[1]{Michael Fire\thanks{mickyfi@bgu.ac.il}}
\affil[1]{Department of Software and Information Systems Engineering, Ben-Gurion University of the Negev, Israel}
    \maketitle

\begin{abstract}
The COVID-19 pandemic outbreak, with its related social distancing and shelter-in-place measures, has dramatically affected ways in which people communicate with each other, forcing people to find new ways to collaborate, study, celebrate special occasions, and meet with family and friends. One of the most popular solutions that have emerged is the use of video conferencing applications to replace face-to-face meetings with virtual meetings.
This resulted in unprecedented growth in the number of video conferencing users.
In this study, we explored privacy issues that may be at risk by attending virtual meetings. We extracted private information from collage images of meeting participants that are publicly posted on the Web. We used image processing, text recognition tools, as well as social network analysis to explore our web crawling curated dataset of over 15,700 collage images, and over 142,000 face images of meeting participants.
We demonstrate that video conference users are facing prevalent security and privacy threats. Our results indicate that it is relatively easy to collect thousands of publicly available images of video conference meetings and extract personal information about the participants, including their face images, age, gender, usernames, and sometimes even full names. 
This type of extracted data can vastly and easily jeopardize people's security and privacy both in the online and real-world, affecting not only adults but also more vulnerable segments of society, such as young children and older adults.
Finally, we show that cross-referencing facial image data with social network data may put participants at additional privacy risks they may not be aware of and that it is possible to identify users that appear in several video conference meetings, thus providing a potential to maliciously aggregate different sources of information about a target individual.

\end{abstract}

%
%

	\providecommand{\keywords}[1]{\textbf{Keywords:} #1}

\keywords{Video Conference, Video Conference Applications, Security and Privacy, Image Processing, Data Science, COVID-19}

\section{Introduction}
\label{sec:int}

The COVID-19 pandemic outbreak has had tremendous impacts on daily life in multiple aspects, ranging from basic hygiene, medicine, health science, economics, politics, and society. In particular, it dramatically affected the ways in which people communicate with their families, friends, and coworkers~\citep{HowtheCo44:online}. 
Social distancing resulted in the search for new ways to collaborate, study, celebrate birthdays, and even meet with parents and grandparents. One of the most popular solutions that have emerged is the use of video conferencing applications, such as Google Meet,\footnote{\url{https://meet.google.com}} Microsoft Teams,\footnote{\url{https://www.microsoft.com/en-us/microsoft-365/microsoft-teams/group-chat-software}} and Zoom.\footnote{\url{https://zoom.us}} During COVID-19 outbreak and accompanying stay-home quarantine strategies, people have started using video conferencing applications to replace face-to-face meetings in schools, workplaces, and social gatherings with virtual meetings.
As a result, the number of video conferencing users along with the number of daily meetings surged sharply~\citep{COVID19O99:online, Zoomadmi15:online}.
Today, there are hundreds of millions of video conferencing meetings that take place daily, encompassing millions of users~\citep{Zoomadmi15:online}.

With the unprecedented growth in video conferencing usage, many security and privacy issues have been unraveled ~\citep{GoogleMe58:online, Zoomsecu72:online}.
These issues range from unencrypted communication for unpaid users~\citep{Zoomunpaid} to vulnerabilities that allow malware execution on participants' devices~\citep{Zoomsecu72:online}.
Moreover, the use of video conferencing applications has raised privacy concerns that may enable uninvited attendees to find ways to join meetings by guessing meetings IDs~\cite{zoomguess} or by simply searching for video conferencing meeting links that were made publicly available (e.g. meeting links published on social media websites~\cite{Zoombom9:online}).

\begin{figure}
  \centering
    \includegraphics[width=1\linewidth]{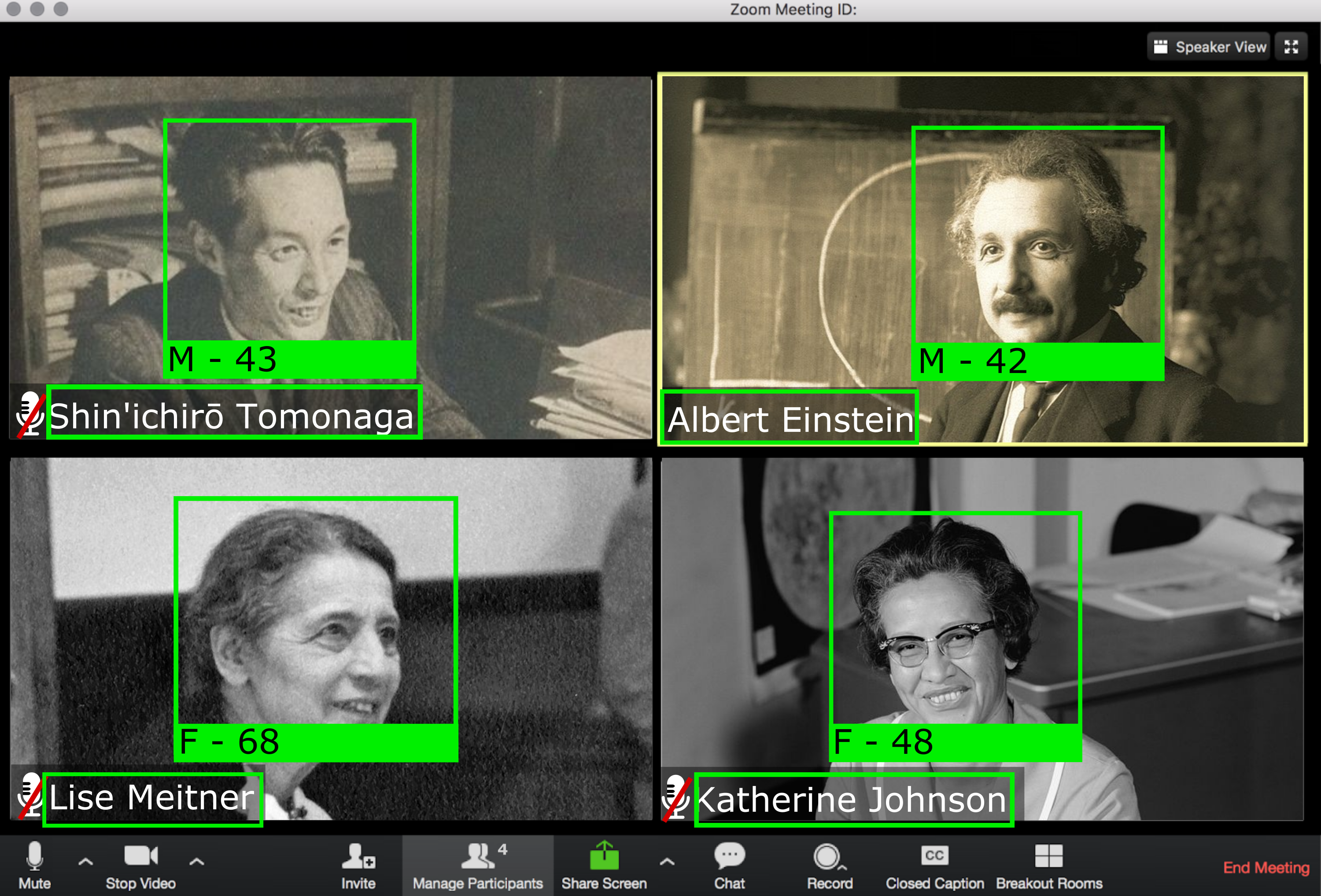}

  \caption{Zoom Image Collage with Detected Information, Along with Extracted Features of Gender, Age, Face, and Username.}
  \label{fig:zoom-detect}

\end{figure}

A malicious user that gains access to video conferencing meetings can collect sensitive and private data on users, such as their names, usernames, images of their faces, samples of their voice, and even exposure to personal data that has been shared as part of the conversations. Moreover, using accessible deepfake tools, a malicious user can attend video conferencing meetings under a false identity. For example, one can use realtime deepfake tools in order to join meetings using celebrity avatars that make him or her look like celebrities, such as Barack Obama or Elon Musk~\citep{ThisOpen93:online}.\footnote{\url{https://github.com/alievk/avatarify}} The participants' personal data can later be used to jeopardize participants' safety in both the virtual and the real-world~\citep{fire2014online}. For example, \citet{acquisti2014face} demonstrated the threat of face recognition can be used to identify individuals both in the online and offline worlds. Namely, they used publicly available images from Facebook to identify students strolling through campus. They also illustrated that it is possible to predict personal and sensitive information from a face, such as the individual’s interests, activities, and even his or her social security number.

In this study, we take an in-depth look into the world of video conferencing by analyzing images taken from thousands of video meetings, which were publicly published online. We curated an image dataset by collecting and analyzing ten-of-thousands images published on social media websites. 
We subsequently analyzed the dataset to extract information from each image regarding the meeting participants, such as their age, gender, username, as well as several other features. 

We compute statistics on video conferencing usage and demonstrate that video conferencing participants encompass a wide range of ages, from young children to older adults (see Figure~\ref{fig:zoom-detect}). We apply deep-learning based image processing algorithms to demonstrate that it is possible to identify the same individual's participation at different meetings by simply using either face recognition or other extracted user features.
The extracted information about users has the potential to be harmfully used to uncover participants' social networks and other privacy related factors (see Figure~\ref{fig:zoom-detect}).
Overall, our study emphasizes the privacy risks that video conferencing participants need to be made aware of. We demonstrate that there is danger not only from a malicious user but also from other users in the meeting that can innocently upload \textit{a single photo}, which can be maliciously be used to affect their and their family's privacy.

The remainder of the paper is organized as follows: In Section~\ref{sec:rw}, we present an overview of relevant studies. In Section~\ref{sec:method}, we describe the datasets, methods, algorithms, and experiments used throughout this study. In Section~\ref{sec:results}, we present our results. In Section~\ref{sec:dis}, we discuss the obtained results. In Section~\ref{sec:lim}, we explain the study's limitations. In Section~\ref{sec:rec}, we recommend mitigations. Lastly, in Section~\ref{sec:con}, we present our conclusions from this study.

\section{Background}
\label{sec:rw}
Over the years, many efforts were put into analyzing user security and privacy in a variety of online platforms, such as online virtual worlds~\cite{zarsky2006privacy}, online trading systems~\cite{tsang2014detecting}, online social networks~\citep{fire2014online} and cryptocurrency platforms~\citep{conti2018survey}. In this study, we focus on privacy and security issues in video conferencing applications, one of the most commonly used one known as Zoom. Such applications have rapidly emerged as cardinal ways of communication as a result of stay-home and social distancing orders due to the COVID-19 pandemic outbreak, and have been increasingly supporting, in addition to work-related and educational activities, also a wide range of social activities, including virtual happy hours, virtual yoga, blind dating, and worship services~\citep{blose2020privacy}.
As a consequence, users of video conferencing platforms have been exposed to multiple security and privacy risks, similar to those related to online social networks. 
\newline
Privacy Risks related to online social networks include the following:
\begin{itemize}
\item \textbf{Cyberbullying.} Cyberbullying (also referred to as Cyber Abuse) refers to bullying and harassment that take place within technological communication platforms, such as chats, mobile devices, and social networks~\cite{slonje2008cyberbullying}. 
Recently, video conferencing applications users have been exposed to a new type of online harassment, termed \textit{Zoom Bombing}, where an uninvited person joins a video chat meeting and interrupts the meeting by sharing inappropriate content~\citep{Zoombom9:online}.

\item \textbf{Information Leakage.} Information leakage refers to the detection and extraction of information that was unintentionally disclosed~\citep{Informat67:online}. Visual data contains implicit information in the background that may reveal interesting, as well as possibly sensitive pieces of information. For instance, \citet{weyand2016planet} was able to detect a photo location exclusively based on image pixels.
\citet{reece2017instagram} demonstrated that it is possible to detect markers of depression in Instagram photos.
\item \textbf{Information Linkage.} Information linkage occurs when an attacker is able to link several pieces of information from at least two separate data sources. For example, an attacker may match user's publicly available details in one online account (e.g. username or profile picture) with other accounts and social network profiles of the same user, in order to uncover additional details about the user, such as full name, home location, and workplace.

\item \textbf{Malware Attacks.} Malware refers to malicious software that is intentionally developed in order to disrupt a computer operation for the purpose of collecting credentials and gaining access to private information. Like many other web platforms, video conferencing applications are also subject to malware attacks. For example,
in parallel to Zoom's dramatic rise in popularity, it has become a popular target for hackers
~\citep{Zoomsecu72:online}. 
Recently, it has been reported that Zero day vulnerability in Zoom allows for remote code execution and launching malware on target computers, \citep{Zerodayv53:online}.

\item \textbf{Phishing Attacks.} Phishing attacks are a form of social engineering that is developed in order to steal user-sensitive and private information by masquerading as a trustworthy third party. Since COVID-19 outbreak, hackers have been increasingly impersonating video conferencing applications, including Zoom, Microsoft Teams, and Google Meet by sending phishing emails and registering over 2,449 Zoom related domains for phishing scams ~\citep{zoomphishing}. 
Moreover, a new type of Zoom related phishing campaign has recently been revealed in which the phishing mail contains a fake Zoom link and includes words like 'termination' to distract users from noticing that the link in the email is not linked to Zoom official website \citep{NewPhish23:online}. 

\item \textbf{Face Recognition Attacks.} Face recognition algorithms are capable of identifying or verifying a person from a digital image or a video source. Identifying a person's face from a video, and cross-referencing it with other datasets might be used to expose personal information about the individual~\cite{ortega2019hiding}. 
\citet{acquisti2014face} demonstrated that it is possible to identify strangers online (on a dating site where individuals protect their identities by using pseudonyms) and offline (in public space), based on photos that are publicly available on social network sites. Zoom, in particular, has enabled until very recently data mining capabilities to secretly display data related to meeting participants by linking participants' names and email addresses with their LinkedIn profiles~\cite{zoomlinkedin}.

\end{itemize}

In addition, users of video conferencing applications are exposed to additional sets of risks:
\begin{itemize}
\item \textbf{Data Breach.} Some video conferencing applications provide recording and cloud storage capabilities for their users.
Recently, the Washington Post~\citep{Zoomvide28:online} reported that thousands of private Zoom videos were publicly accessible online on either Zoom's cloud storage, or on other external storage services.
Washington Post reporters watched several of these videos and discovered that these videos contained sensitive and private information, including people’s names and phone numbers, private company and financial statements in small-business meetings, as well as elementary school classes, in which children’s faces, voices, and personal details were exposed. Even users that used a password to protect their videos on Zoom cloud were exposed to these same risk, as presented by a proof of concept example that used brute force to bypass the password protected videos.\footnote{https://github.com/markbuffalo/zoombo}

\item \textbf{Fake Avatars.} Fake avatar is a new type of identity fraud, similar to fake profiles in online social networks~\cite{fire2014friend}. Video conferencing application users can be victims of this new type of identity fraud, in which other participants in a meeting can pretend to be a person who they are not. For example, the code project Avatarify\footnote{\url{https://github.com/alievk/avatarify}} can be used to create photorealistic avatars for video-conferencing, that may mislead meeting participants to think they are meeting with a celebrity~\cite{zoomdeepfake}. Similar to other deepfake based applications, this type of technology might be used for fraudulent activities. For example, in 2019, it is believed that fraudsters used deepfake voice software to mimic a CEO’s voice in order to steal money from a company~\cite{stupp2019fraudsters}.

\item \textbf{Zoombombing Campaigns.} Zoombombing (also referred to as Zoom Raiding) is a practice that refers to unwelcome stranger participants joining a meeting, and disrupting video calls with offensive language and imagery. This type of behavior has recently become a dangerous method for coordinated harassment and hate speech~\cite{Zoomriding}. 
For example, recent research performed by the New York Times uncovered that thousands of people gathered online to organize Zoom harassment campaigns, share meeting passwords and plan for sowing chaos in public and private meetings~\cite{Zoomriding}. 
Also, in the UK, there is an ongoing investigation of 120 Zoombombing cases of Zoom video calls that were hijacked by people displaying images of child abuse~\citep{Morethan8:online}.

\end{itemize}

\section{Methods and Experiments}
\label{sec:method}
The primary goal of this study was to explore privacy aspects that may be at risk by attending virtual conference meetings, a practice that has become very common following the COVID-19 outbreak and related stay-home lockdown and social distancing measures.

To achieve this goal we performed the following steps:

\subsection{Curating an Image Dataset of Video Conferencing Collages}
\label{sec:dataset}
\begin{figure}
  \centering
    \includegraphics[width=1\linewidth]{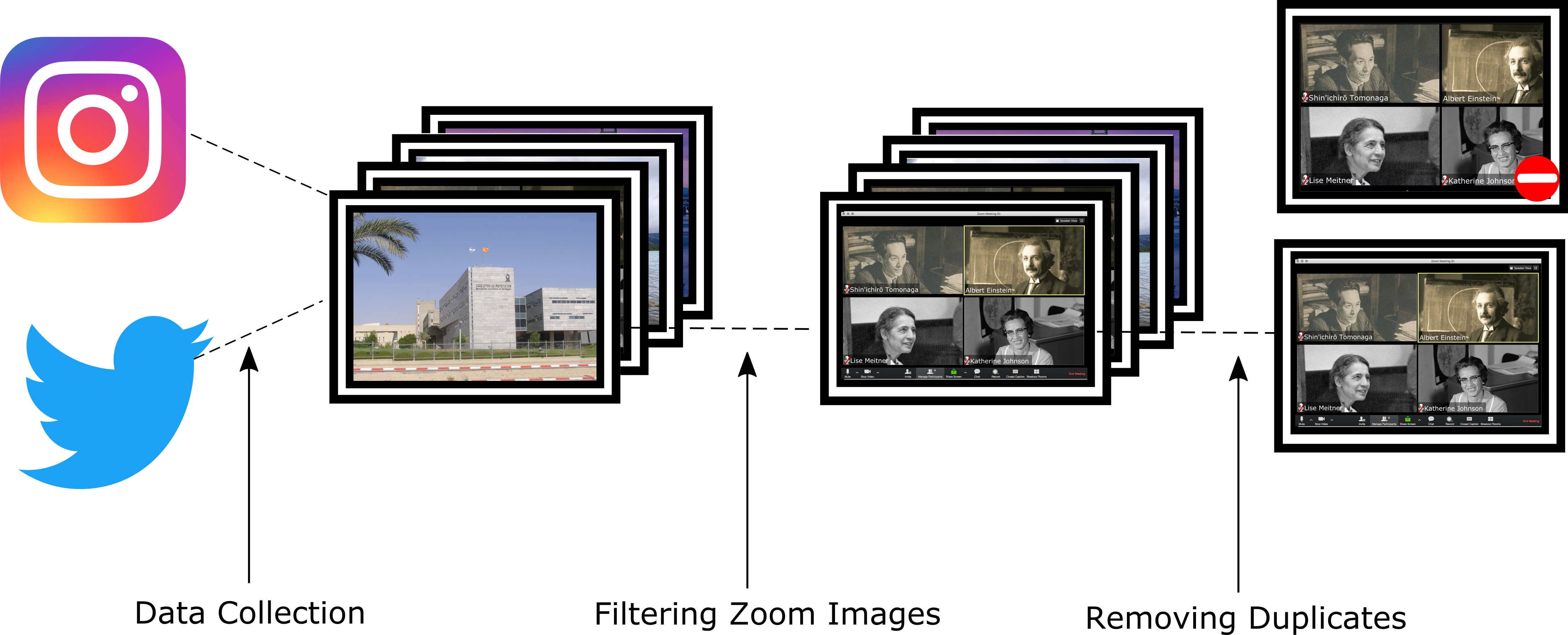}

  \caption{Dataset Generation Process.}
  \label{fig:prepro}

\end{figure}

We curated an image dataset that contains images from thousands of video conference meetings by performing the following three steps (also illustrated in Figure~\ref{fig:prepro}): 

First, to collect images from video conference meetings, we utilized web crawlers that collect data from Twitter and Instagram using online accessible Twitter and Instagram scraper tools.\footnote{\url{https://developer.twitter.com/en/docs}}\footnote{\url{https://github.com/arc298/instagram-scraper}}
We were specifically interested in collecting images that were likely to be taken in video conference meetings, thus we set the web crawlers to search for tweets that contain terms or hashtags from a predefined set of meeting related target terms (e.g.``zoom school'' and ``\#zoommeeting'').\footnote{Our twitter crawler downloaded tweets that contained the following terms and hashtags: ``zoom birthday,'' ``zoom happyhour,'' ``zoom school,'' ``zoom party,'' ``\#happyhour,'' ``\#Zoom,''``\#zoommeeting,'' ``\#zoomus,'' and ``@Zoom\_US.'' The Instagram crawler downloaded posts containing the following hashtags: ``\#zoommeeting,'' ``\#zoomparty,'' and ``\#zoombirthday.'} 
Using this method we were able to collect 89,305 Zoom related tweets and 90,395 Zoom related Instagram posts (a total of 179,700 meeting related posts).

Second, we filtered out tweets and Instagram posts that did not contain an image and extracted the images from all remaining tweets and posts (a total of 26,408 images from Twitter and 78,435 images from Instagram were collected). 
We then utilized Fastai framework~\citep{howard2018fastai} to construct an image classifier that identifies which of the images contain a Zoom video meeting collage of participants. 
 
To train the classifier, we manually labeled 5,271 images, out of which 1,505 contained a collage of participants (positive examples), and 3,766 were labeled as other types of images (negative examples). Using the labeled images, we applied ResNet-50~\cite{he2016deep} based transfer learning to train our model.
The trained classifier achieved an accuracy of 0.969, a true-positive rate of 0.935, and a false-positive rate of 0.016 on a test set of 1,054 images.
This process resulted in a dataset consisting of 16,133 Zoom collage images.

To remove duplicates and similar images (i.e same image with minor adjustments or crops),
we utilized dhash~\citep{KindofLi17:online} to compute the Hamming distance between all pairs of images. We removed all images with distance equal or smaller than an empirically set threshold 1.2.\footnote{The selected threshold of 1.2 was chosen, by manually testing and evaluating several threshold values, searching for minimum false positives.}
Moreover, to remove additional similar images, we computed the Euclidean and cosine distances between all pairs of images using each image embedding generated by using the Zoom image classifier last layer representation. Then, we removed all the images that their distance was equal or lower of 0.0035 and 25 for cosine and euclidean distance respectively.\footnote{The selected threshold values of 0.0035 and 25 were empirically chosen, by testing and manually evaluating several threshold values.}
Finally, we remained with an image dataset that consists of 15,709 collage images automatically collected from Twitter and Instagram.

\subsection{Feature Extraction by Image Processing of Collages}
\label{sec:features}
To extract structure data from collage image dataset, we performed the following steps on each image in the dataset:
\begin{enumerate}
    \item \textbf{Faces Recognition.} Zoom collage images typically contain multiple faces. We utilized two face recognition tools to detect each of the faces in a collage. The first tool we used was a pre-trained model~\citep{ipazcmtc85:online},  based on MTCNN~\citep{zhang2016joint}. 
    The second tool we utilized was Microsoft Azure Face API for face recognition~\citep{FacialRe52:online}. 
    Each of the two tools extract the bounding boxes of the faces in each collage image in our dataset. Then, the results of the two face recognition tools are combined in the following manner: We keep the faces that are uniquely identified by each tool, and in cases where bounding boxes of two faces intersect, we consider them as the same face (the bounding box from the first model is used). To estimate recall values, 
    we manually evaluated faces detection recall rates in 100 random selected Zoom collage images. Our face detection approach achieved a recall of 80\% (i.e detection of 80\% faces of the available data).
    
    \item \textbf{Face Embedding.} By using a dlib based tool~\citep{dlib09,ageitgey33:online}, each detected face in our dataset, we generated a 128-dimension numerical vector representation, which represents the features extracted from the face. 
    
    \item \textbf{Age Detection.} For each detected face, we estimated the age of the person, by applying two separate models for inferring the age of an individual based on his or her face image, a pre-trained model for age detection~\citep{yu4uageg58:online}\footnote{The code was refactored into a library instead of a CLI tool. The modified code is available on  \href{https://github.com/Kagandi/age-estimation-pytorch}{GitHub}.}
    and the package of Microsoft Azure Face API~\citep{FacialRe52:online}. 
   We then define the estimated age as the average value of two predicted ages. 
   We binned the average predicted ages into four categories~\citep{Age9:online}:
\[
    f(x)= 
\begin{cases}
   child,& \text{if } x\leq12\\
   adolescent,& \text{if } x> 12 \:and\: x\leq17\\
   adult,& \text{if } x> 17 \:and \: x<65\\
    older\: adult,              & \text{otherwise}
\end{cases}
\]
   
   \item \textbf{Gender Detection.} For each face detected by Microsoft Azure Face API~\citep{FacialRe52:online}, we also utilized the API to detect the user's gender. By manually inspecting 100 randomly selected Zoom collage images, we evaluated that Microsoft Azure Face API detects gender on our curated dataset with a recall of 98.9\%.

      \begin{figure}
  \centering
    \includegraphics[width=0.9\linewidth]{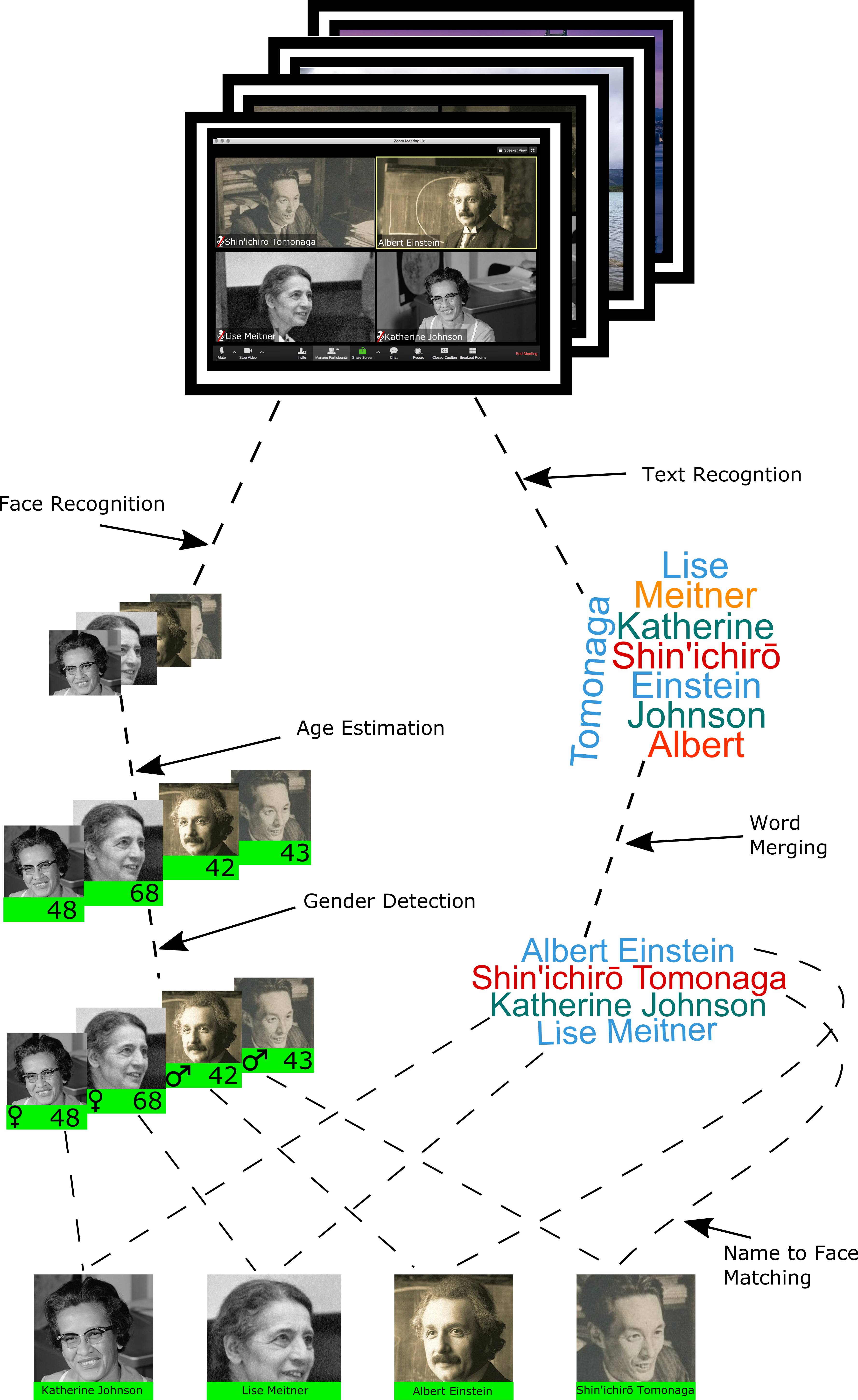}

  \caption{Data Extraction Process}
  \label{fig:extract}

\end{figure}

   \item \textbf{Username Recognition.} To identify participants' usernames, 
   we utilized a publicly available scene text recognition library.\footnote{\url{https://github.com/gtsoukas/scene\_text}}
   The library uses a text detection approach that is based on  EAST~\citep{zhou2017east} for detecting text location and MORAN~\citep{luo2019multi} for subsequently recognizing the composing characters.
   To filter out words that are not username candidates, we used a list of Zoom meeting-related words.\footnote{The list consists of the following words: ``help,'' ``view,'' ``meeting,'' ``edit,'' ``participant,'' ``speaker,'' ``security,'' ``screen,'' ``record,'' ``video,'' ``stop,'' ``share,'' ``manage,'' ``view,'' ``exit,'' ``full,'' ``chat,'' ``end,'' ``reactions,'' ``recording,'' ``mute,'' ``zoom,'' ``participants,'' ``from,'' ``recording,'' ``window,'' ``search,'' ``invite,'' ``leave,'' ``unmute,'' ``option,'' ``delete,'' ``raise,'' ``hand,'' ``new,'' ``type,'' ``message,'' and ``here.''}
    We used spaCy lemmatization~\citep{spacy2} to filter out all dictionary words, aiming at removing words that are not names. 
  One of the issues with the text recognition model is that it detects only single words. However, in many cases usernames are a combination of multiple words, such as first name, middle name, and last name. To combine single detected words into the associated full usernames we applied a custom heuristic approach based on the unique image structure of Zoom collages.
   We noticed that it is seldom to find two words located very close one to another that are not a part of the same name. Based on this observation, we developed a simple heuristic to join recognized words that are closely-located in image space into a single username. 
   Namely, for each identified word in each image collage, we searched for another word that is the nearest to its top right point.
    If the nearest word was in a euclidean distance smaller or equal to 10,\footnote{The selected threshold of 10 was chosen, by manually testing and evaluating several threshold values.} we combined both words. To construct usernames of more than two words, we repeated this process repeatedly until no words were further combined.
    We evaluated this heuristic on 100 randomly selected Zoom collage images and found that it combines 97.6\% of the usernames longer than a single word into the correct usernames. 
    To evaluate the username recognition (scene text detection that is followed by word merging) we compared the actual username with the detected username in 100 randomly selected Zoom collage images. The method was able to detect 63.4\% of the usernames correctly.\footnote{We used a strict measurement where even if only one character is different between the actual username and detect one it will not be considered as correct detection.}

\end{enumerate}

\subsection{Linking Personal User Data to Social Network Data}

We explored whether video conference participants are vulnerable to information leakage attacks, such as linkage to social networks. To demonstrate that this type of attack is indeed possible using video conference data, we reconstructed participants' links and social networks by matching users that participated in multiple meetings using the following methods:

(a) \textit{Cross Referencing Usernames}: We match users across video conference meetings by explicitly using participants' usernames. This method considerably reduces the search space of  of Zoom participant identities.
Moreover, in some cases, participants have unique usernames, which greatly assists in finding these same individuals in multiple meetings 

(b) \textit{Cross Referencing using Face Embedding}: We match users across different meetings by comparing the detected participants' faces images. Namely, we utilized the extracted face embeddings (see Section~\ref{sec:features}), and calculated the euclidean distance between all pairs of detected faces. We consider two face images to be images of the same user if the euclidean distance between their vector representation is smaller than an empirically chosen threshold.\footnote{To determine that two face images are of the same person, we empirically chose a distance threshold between the vector representation of two faces to be less or equal 0.3, which is a lower than the default distance value of 0.6 used by the \textit{Face Recognition} python library~\citet{ageitgey33:online}. We chose a relatively strict threshold since many of the faces extracted from Zoom collage images suffer from low resolution, thus resulting in many false matches of pairs of faces.}

Using the cross-referenced data, we constructed a social graph $G:=<V,E>$ of the participants, where $V$ is the set of video conference users, and $E$ is a set of links between users that according to the collage image dataset participants in a meeting together.

\begin{figure}
  \centering
    \includegraphics[width=0.75\linewidth]{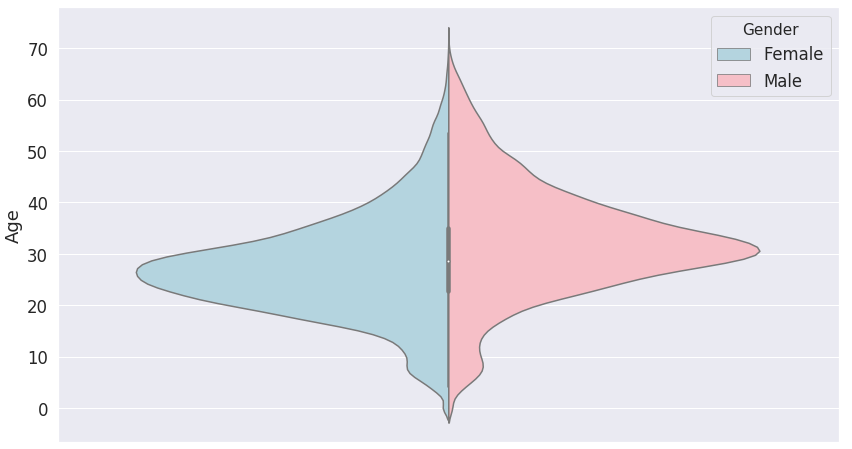}

  \caption{Zoom users gender and age distribution.}
  \label{fig:gender}

\end{figure}

In order to demonstrate that video conference meeting participants are susceptible to information leakage attacks, we tested whether by using the participants' usernames and facial images, it is possible to match video conference users with their online social networks' profiles. To achieve this goal, we manually searched for usernames on social media sites and compared them to a video conference username with the same face. In addition, after matching a video conference user with his or her online social network profile, we explore whether it is possible to gather additional data about his or her friends to reverse search and match the identified friends profiles with their video conference profiles.

\section{Results}
\label{sec:results}

To analyze privacy threats in video conference collage images (see Section~\ref{sec:method}), we collected 15,783 Zoom collage images that were publicly posted on Twitter and Instagram.
We explored the dataset of collected Zoom collages and found an average of 9.04 participants in a video conference meeting collage. From the crawled Zoom collage images, we extracted a total of 142,001 face images for analysis. 
From an age perspective (see Figure \ref{fig:gender}), the average estimated age of the participants is 29.23 and the median is 28.56.
By examining age distributions, we observed that 87.49\% of the participants are adults, 6.18\% children, 6.23\% adolescents, and only 0.09\% of participants are older adults.
Additionally, the gender prediction algorithms (see Section~\ref{sec:method}) were able to identify 29,048, and 50,221 males and females respectively. 

\begin{wrapfigure}{r}{0.5\textwidth}
  \centering
    \includegraphics[width=1\linewidth]{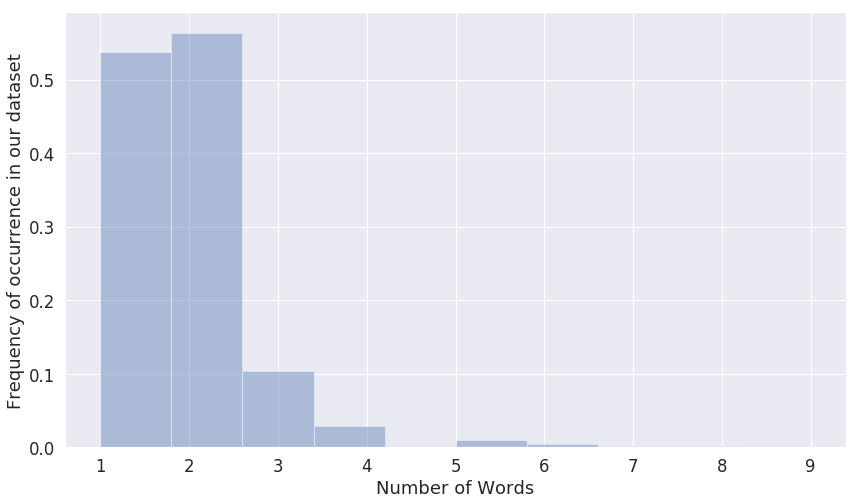}

  \caption{Distribution of the Number of Words Composing a Zoom Username }
  \label{fig:words-dist}

\end{wrapfigure}

We were interested in how discrete users are about their real world identities, as reflected in the choice of their usernames. 
We thus extracted usernames from the Zoom image collages, 
and found 85,616 distinct usernames, out of which 48,818 (57\% of the usernames) consist of more than a single word (see Figure \ref{fig:words-dist}), suggesting possibly more disclosed information about the user.
By manually inspecting these 48,818 multi-word usernames we observed that many of these words represent distinct names that can be utilized to match a user's social network profile.
In fact, 2,522 of these usernames appeared in several collage images from different meetings. 
Moreover, some users use the name of their phone model as a username, thus not disclosing personal identity by the choice of username. In fact, we found that in our dataset the username ``iPhone'' was the most popular single word username (395 appearances).

\begin{figure}
  \centering
    \includegraphics[width=1\linewidth]{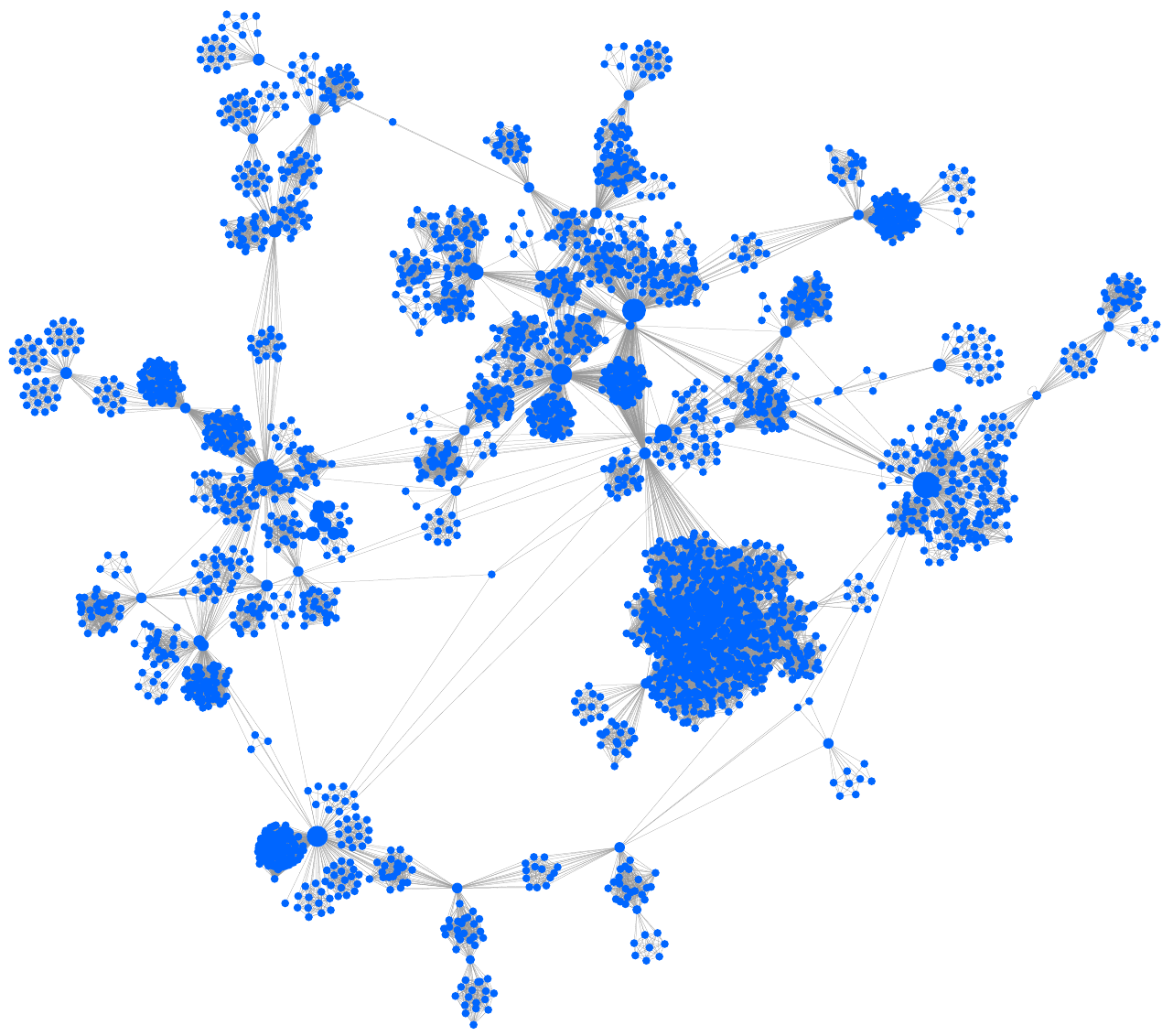}

  \caption{The Largest Component Observed in the Social Network of Zoom Users in Our Dataset. 
  The network was constructed by matching meeting participants' faces and usernames across meetings as explained in Methods (see Section \ref{sec:method}). Each node in the network represents a Zoom meeting participant, and each edge between users in the network represents that both users participated in at least in one video conference meeting. }
  \label{fig:network}

\end{figure}

By analysing the participants facial images, we identified 1,153 faces that likely appeared in several different Zoom meeting images.
Using the cross-referenced images we constructed a large-scale social network of Zoom users with 16,842 nodes (participants) and 197,765 edges (participation in joint meetings).
Each node in the network represents a single meeting participant, and each edge between a pair of participants in the network represents that they jointly participated in at least in one video conference meeting. Users may jointly participate in meetings from different worlds of content.
The network consists of distinct 345 connected components.
On average, each separate component consisted of 48.8 participant nodes and 573.2 joint meeting edges respectively. The largest component consisted of 3,066 nodes and 55,035 edges (see Figure~\ref{fig:network}).
By inspection of randomly selected participants, we were able to manually locate their personal social network profiles. We also observed networks where all participants were co-workers.
Finally, we found that some users are more privacy aware than others, using smiley emojis above their faces to protect their privacy.

\section{Discussion}
\label{sec:dis}
Our results point to several important findings:
 
First, we demonstrate how easy it is to collect personal data from an abundance of Zoom collage images that are publicly posted on the Web.
Using Instagram and Twitter crawlers, we were able to collect over 15,000 collage images (see Section~\ref{sec:dataset}), all taken during Zoom video conference meetings. From those Zoom collages, using face recognition algorithms, we were able to extract a dataset of over 140,000 faces and over 85,000 distinct usernames. These results indicate that many images taken from video conference meetings are already publicly available online, and relatively easy to collect. Moreover, the faces and usernames collected in this type of process can be used to construct a facial image dataset, which contains personal details about meeting participants, including facial characteristics, age, gender, usernames, and sometimes even full names. 
This type of facial image dataset can vastly and easily jeopardize people's security and privacy both in the online and real-world. Furthermore, in this study, in order to demonstrate this point, we generated our dataset by collecting publicly available data only. However, malicious users, such as zoombombers, can potentially actively harvest data from video conference meetings and construct more precise facial datasets of video conference participants' data without their consent and knowledge.

Second, the collected dataset provides us with a demographic glimpse into the world of video conference meetings. By analyzing age and gender features of over 140,000 participants (Figure~\ref{fig:gender}), we observed that today's video conferencing is widely used by various segments of the population.
While the vast majority (87.49\%) of meeting participants were adults, as might be expected considering that due to COVID-19 many companies have been encouraging employees to work from home~\cite{wfh2020}, the age span is in fact from children to older adults. This result highlights the risk that privacy and security issues may affect not only adults but also more vulnerable segments of the population, such as young children and older adults.

Third, we showed that cross-referencing facial image data with social network data may put participants at additional privacy risks they may not be aware of.
In fact, as one example, we explored a meeting that contained a group of adults. By performing an in-depth analysis of these images and comparing the participants' usernames and facial images with data available from online social networks, we were able to conclude that the meeting participants worked in the same company and that we could additionally infer social links among the meeting participants. 
This illustrates that not only individuals' privacy is at risk from data exposed on video conference meetings, but also the privacy and security of organizations. 
For example, by inspecting the social network of participants in an organization, it possible to expose a variety of private details regarding the organization itself~\cite{fire2016organization}. Thus, an attacker can potentially utilize the organization's social network in order to attack the organization~\cite{paradise2014anti}. 

Fourth, we demonstrated that by using meeting participants' facial characteristics and usernames, it is possible to identify users that appear in several video conference meetings, thus to collect and aggregate their data. For example, as depicted in Figure~\ref{fig:network}, it is possible to construct an individual's social network by collecting data from several meetings he or she participated in. 
Moreover, we demonstrate that it is possible to use data collected from video conference meetings along with linked data collected in other video meetings with other groups, such as online social networks, in order to perform a linkage attack on target individuals. This can result in jeopardizing the target individual's privacy by using different meetings to discover different types of connections.
For instance, one may aggregate information regarding job ,personal, and political related social connections.

Finally, during this study, we observed that some meeting participants were more successful than others in protecting their privacy. Such privacy protection practices included the use of generic usernames, replacing their faces with smiley images, etc. Interestingly, even for such privacy aware individuals, we realized that it is possible to identify the same individuals across different meetings by using other means, such as identifying unique backgrounds, another potential privacy risk that users may be unaware of.

\section{Research Limitations}
\label{sec:lim}
It is worth noting that the methods used in this study are prone to several limitations:
First, we collected only publicly available images from Twitter and Instagram postings. Therefore, our collage image dataset is partial and does not contain data from meetings that were not publicly published online. 
Additionally, we analyzed the images using state-of-the-art facial recognition tools. Nevertheless, these tools and algorithms have been reported to be biased in some cases ~\citep{garvie2016facial}.
To overcome such types of face recognition issues we combined the results of two facial recognition models (see Section \ref{sec:features}). Also, to diversify the datasets, we collected data from multiple sources, Twitter, and Instagram.
Moreover, the quality of the Zoom collage images in our dataset is not consistent, affecting our extracted data. Some images are screenshots and others taken by smartphones. Additionally, there are differences in the image resolution, which directly affects the number of pixels available for analysis of each face and username. The lower the image quality the harder it to extract accurate data.
We also observed that some actions can improve the method general accuracy, including image alignment, face verification, and training a model for splitting a collage into rectangles may improve performance of name to face matching as well as face detection.
Lastly, retraining the models on Zoom data may significantly improve the accuracy of the models used in Section \ref{sec:features}.

\section{Recommendations}
\label{sec:rec}
As we have demonstrated throughout this study, video conference users are facing prevalent and varied security and privacy threats.
In this section, we provide several easy-to-apply methods which can assist video conference participants to improve their security and privacy:
\begin{itemize}
    \item \textbf{Avoid Video Streaming Whenever Possible.} As we have shown in this study, images taken from video conference meetings can jeopardize individual security and privacy in multiple ways. Even in cases of private meetings, it is sufficient for a single collage image to be taken by one of the participants in order to jeopardize the privacy of all associated meeting participants. Therefore, whenever possible, we strongly recommend video conference users to not share videos of meetings.

    \item \textbf{Avoid Uploading Virtual Meeting Photos and Videos onto Social Media.}
    Avoid uploading videos and collage photos of video conference meetings onto social media.

    \item \textbf{Use Generic Pseudonames for Video Conferencing.} As demonstrated in this study, using a full name or a unique username for video conferencing can easily expose an individual to information linkage attacks. Moreover, using a generic name, such as ``iPad'' or ``iPhone'' makes it more challenging to cross-reference a user's identity with other datasets. Therefore, we strongly recommend to video conference participants not to use their real name and advise using a generic pseudonym instead.
    
    \item \textbf{Use Generic Background Images as Virtual Backgrounds}. While video streaming, using a virtual background is highly recommended. Using a real background can compromise a user's security and privacy. For example, items in the background of an individual's room can expose his or her name and even geographic location. Additionally, not using a virtual background or using a unique background image may help malicious users in fingerprinting a user account across several meetings.
    
    \item \textbf{Use Anti-Facial Recognition Accessories.} In recent years, some companies have started selling anti-facial recognition cloth and makeup to protect the privacy of users~\citet{Theseclo16:online}. Some accessories can be homemade. For example, \citet{sharif2016accessorize} presented a method that is using a home printer to create skin for glasses that disrupt facial recognition algorithms.
    Moreover, we observed that in many cases, we were not able to detect the faces of people wearing even a simple party mask.

    \item \textbf{Organizations Should Inform Employees on Video Conference Privacy Risks.} Many organizations, especially during times of pandemic, suggest that their employees work from home and conduct meetings using video conference applications. This results in a new set of security and privacy threats to organizations that not all organizations are aware of. In this study, we demonstrated that sharing data from video conference meetings can compromise organizations' security and privacy. Therefore, we recommend organizations be vigilant to the new risks, which are raised with the increased usage of this technology. Moreover, we recommend organizations to update their policies on what employees can share and not share while using video conference applications.

    \item \textbf{Monitor Children’s Video Conferencing Activity.} As we observed in our study, children are active participants in video conference meetings. While video conference may seem as a safe and harmless place to be, they are at risk of numerous threats ranging from malware to zoombombing as described in Section \ref{sec:rw}. Video conference is not different from regular online activity and parents should be actively involved in monitoring their children's activity the same way.
    To reduce risks, parents should explain to their kids not to accept a request to join a meeting from someone they do not know in real life, and also remove personal details from the child's account and adjust the privacy settings.

    \item \textbf{Video Conference Operators Should Add and Support Privacy Mode.} In the past couple of years, researchers presented methods that can disrupt facial recognition. 
    For instance,~\citet{wilber2016can} showed that adding Gaussian noise to an image can disrupt facial recognition while keeping the face still recognizable for humans. 
    Even without the help of video conference operators in some applications, users may opt to use filters in order to hide their appearance and to avoid automatic facial recognition.

\end{itemize}

\section{Conclusions}
\label{sec:con}

In 2020, the COVID-19 pandemic drastically and globally changed the way people communicate, immensely increasing communication via video conferencing.
As a result, novel privacy and security issues have emerged, endangering hundreds of millions of video conference participants.
In this study, we dive into the privacy issue of video conference applications by analysing over 15,000 video conference images of over 140,000 meeting participants. 
From these images, we extracted multiple private information features including gender, age, and real name.
We demonstrate that this information can be utilized to uncover additional details about video conference participants.

We discovered that cross-referencing participants in multiple conference meetings can uncover the participant's social network and subsequently may lead to compromising his/her or other participants' privacy and social accounts. That said, we offer several recommendations in which users can protect themselves, for instance by using generic usernames and background, wearing small masks, or using a filter. We observed that some users pasted emojis on their faces to protect their privacy, and we found this helpful for protecting their privacy.
In the current global reality of social distancing, we must be sensitive to online privacy issues that accompany changes in our lifestyle as society is pushed towards a more virtual world.

\section{Data availability}
Due to the private nature of the data, an anonymized version of the constructed social network is available only upon request from the corresponding author.

\bibliographystyle{unsrtnat}
\bibliography{sample-bibliography}

\end{document}